\documentclass[10pt,twocolumn]{article}
\usepackage{amsmath}
\usepackage{amssymb}
\usepackage{graphicx}
\usepackage{fullpage}
\usepackage{float}
\usepackage{algorithm}
\usepackage{algorithmic}
\usepackage{hyperref}
\usepackage{authblk}
\usepackage{breqn}

\author{Vincent Delos$^*$ and Denis Teissandier$^{**}$}
\date{
    University of Bordeaux \\
    CNRS, National Center for French Research \\
    I2M, UMR 5295 \\
    Talence, F-33400, France \\
    $^*$E-mail: \url{v.delos@i2m.u-bordeaux1.fr} \\
    $^{**}$E-mail: \url{d.teissandier@i2m.u-bordeaux1.fr}\\
}

\title{Minkowski sum of $\mathcal{HV}$-polytopes in $\mathbb{R}^n$}

\begin{document}

\maketitle

\begin{abstract}
Minkowski sums cover a wide range of applications in many different fields like algebra, morphing, robotics, mechanical CAD/CAM systems ... This paper deals with sums of polytopes in a $n$ dimensional space provided that both $\mathcal{H}$-representation and $\mathcal{V}$-representation are available i.e. the polytopes are described by both their half-spaces and vertices. The first method uses the polytope normal fans and relies on the ability to intersect dual polyhedral cones. Then we introduce another way of considering Minkowski sums of polytopes based on the primal polyhedral cones attached to each vertex.

{\bf keywords:} Computational Geometry, Convex Polytope, Minkowski Sum, Normal Fan, Polyhedrical Cone.
\end{abstract}

%
%
\section{Introduction}
In mechanical design, tolerancing analysis consists in simulating the geometric variations due to the manufacturing process. A common way to simulate the variations of an over-constrained mechanical system is to manipulate sets of constraints in $\mathbb{R}^6$, to limit the 6 degrees of freedom (3 translations and 3 rotations), see \cite{Homri2013}. In order to compute the cumulative stack-up of variations we need to calculate the Minkowski sums \cite{Srinivasan1993} and intersections of sets of contraints modelled by polytopes in $\mathbb{R}^6$. Two algorithms were developped in this direction in \cite{Teissandier-hal-00635842} and \cite{Wu2003} but only in $\mathbb{R}^3$. In \cite{DelosGTA} and \cite{Fukuda20041261}, Delos and Fukuda introduce other methods summing polytopes in $\mathbb{R}^n$ but they only work with the polytopes $\mathcal{V}$-description and in tolerancing analysis, the polytopes are first defined by half-spaces. We can take advantage of this important property in order to set up the following algorithm. Moreover, making use of the $\mathcal{H}$-representation can speed up the algorithm and is a key element in computing intersections further on. And finally, as we are in small dimensions, computing the double description is not a problem as stated in \cite{Fukuda20041261} we can find cases where: \textit{``the number of facets of the convex hull of a set of k points in Euclidean n-space can be of order $\mathcal{O}(k^{\lfloor n/2 \rfloor})$ even when n is considered fixed''}.

The goal of this paper is to describe two ways of computing Minkowski sums of polytopes, whether we choose to work in the primal or dual spaces. The dual space approach has come to an algorithm implemented and tested in C++ while the other one is still under investigation. We finally introduce some promising perspectives to reach the objective of having a stable algorithm in the field of tolerancing analysis.
%
%
\section{Basic properties}

\subsection{Minkowski sums}

Given two sets $A$ and $B$, let $C$ be the Minkowski sum of $A$ and $B$
\begin{equation}
  C = A+ B = \{ c \in \mathbb{R}^n, \exists a \in A, \exists b \in B / c = a+b \}.
\end{equation}

\begin{figure*}
  \begin{center}
    \includegraphics[scale=0.33]{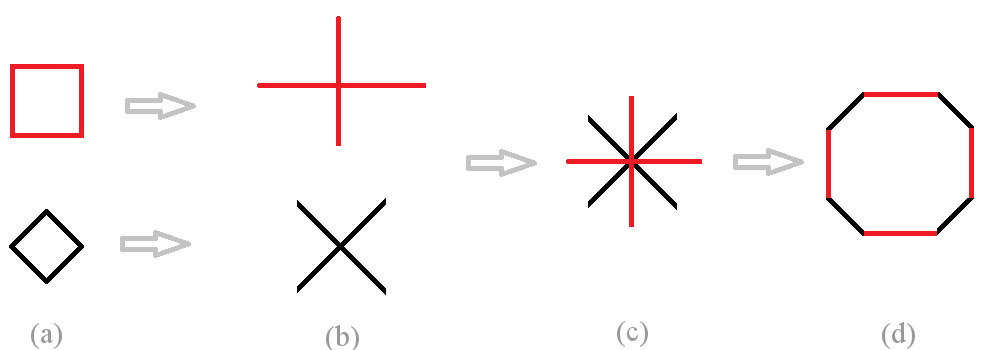}
    \caption{\fontsize{8}{9.6}{Sums of two polytopes step by step: a) polygons, b) normal fans, c) common refinement, d) sum.}}
    \label{com_raf}
  \end{center}
\end{figure*}

\subsection{Polytopes}

A \emph{polytope} is defined as the convex hull of a finite set of points, called the $\mathcal{V}$-representation, or as the bounded intersection of a finite number of half-spaces, called the $\mathcal{H}$-representation. The Minkowski-Weyl theorem states that both definitions are equivalent.

\subsection{Normal fans}

For each vertex $v$ of a polytope, we define the set $E$ of its edges oriented towards its neighbours. With $E=\{e_1, ..., e_l\}$ we build a polyhedral cone $C(v)$ named the primal cone of $v$: 

\begin{equation}
C(v) =\{u_1e_1+...+u_le_l, \forall u_j \ge 0\}.
\end{equation}

We can note that a polytope can be written as the intersection of all the primal cones attached to its vertices. Let $A$ be a $\mathbb{R}^n$-polytope and $\mathcal{V}_A$ the list of its vertices such as $card(\mathcal{V}_A)=m$:

\begin{equation}
A = \displaystyle{\bigcap_{i=1}^{m} C(v_i) }.
\end{equation}

For each vertex $v$ of a polytope, we define the set $N$ of the outer normals of its corresponding facets. With $N=\{n_1, ..., n_k\}$ we build a polyhedral cone $C_D(v)$ named the dual cone of $v$: 

\begin{equation}
C_D(v) = \{t_1n_1+...+t_kn_k, \forall t_i \ge 0\}.
\end{equation}

It is also the set of hyperplane outer normals which find their maximum on this specific vertex:

\begin{equation}
C_D(v) = \{ y \in \mathbb{R}^n : {\langle v, y \rangle} = sup_{x \in A} {\langle x, y \rangle} \}.
\end{equation}

$N(A)$ is defined as the set of all the dual cones of a given polytope, it forms a partition of the whole space $\mathbb{R}^n$ which is called the \textit{normal fan} of the polytope.

\section{Dual algorithm}

As stated in \cite{Weibel3883} by Weibel \textit{``The normal fan of polytopes contains all of their combinatorial organization. It is therefore enough to compute the normal fan of a Minkowski sum to have its combinatorial properties. We can then easily deduce the polytope itself by combining these informations with the summand polytopes. We know that the normal fan of a Minkowski sum is the common refinement of the normal fans of its summands.''} This is why we developped such an approach in a previous article named \textit{Algorithm to calculate the Minkowski sums of 3-polytopes based on normal fans} in $\mathbb{R}^n$. To extend the results in $\mathbb{R}^n$ we need some theoritical results.

\subsection{Main properties}

Ziegler and Gritzmann give the normal fan of the sum of two $\mathbb{R}^n$-polytopes in \cite{ziegler1995lectures} and \cite{Gritzmann1993}. Let $A$ and $B$ with their respective lists of vertices $\mathcal{V}_A$ and $\mathcal{V}_B$:

\begin{equation}
  N(A+B) = N(A) \wedge N(B).
\label{raff_commun_eq_2}
\end{equation}

where $N(A) \wedge N(B) = \{ C_D(a_i) \cap C_D(b_j) : C_D(a_i) \in N(A) ~\forall a_i \in \mathcal{V}_A, C_D(b_j) \in N(B) ~\forall b_j \in \mathcal{V}_B \}$ is called the common refinement. So it is clear that computing the sum of two polytopes can be performed by intersecting polyhedral cones. This is illustrated in Fig. \ref{com_raf}.

In the following we will emphasize on how finding the Minkowski vertices i.e. the vertices of the two operands sum.

\subsection{Minkowski vertices}

Let $A$, $B$ and $C$ be three $\mathbb{R}^n$-polytopes such as $C = A + B$, let $\mathcal{V}_A$, $\mathcal{V}_B$ and $\mathcal{V}_C$ be their lists of vertices. Let $c \in \mathcal{V}_C$, from \cite{Fukuda20041261} we know that the vertex $c$ can only be the sum of a face from $A$ and a face from $B$. For reasons of dimension, $c$ is necessarily the sum of two vertices $a \in \mathcal{V}_A$ and $b \in \mathcal{V}_B$. Let's characterize now the dual cone of $c$.

\begin{equation}
  c=(a+b) \in \mathcal{V}_C \Rightarrow  C_D(c) = C_D(a) \cap C_D(b).
\label{cone_sommet}
\end{equation}

Let $ u \in C_D(a) \cap C_D(b) $ then $ {\langle u, a \rangle} = sup_{x \in A} {\langle u, x \rangle} $ and $ {\langle u, b \rangle} = sup_{y \in B} {\langle u, y \rangle} $. So $ {\langle u, a+b \rangle} = sup_{x \in A, y \in B} {\langle u, x+y \rangle} $ which means that $ {\langle u, c \rangle} = sup_{z \in C} {\langle u, z \rangle} $ i.e. $ u \in C_D(c)$.

On the other side let $ u \in C_D(c) $, by definition $ {\langle u, c \rangle} = sup_{z \in C} {\langle u, z \rangle} $. If we decompose $c$ in a sum of vertices $a$ and $b$ we get $ {\langle u, a \rangle} + {\langle u, b \rangle} = sup_{x \in A} {\langle u, x \rangle} + sup_{y \in B} {\langle u, y \rangle} $, but we know that in general $ {\langle u, a \rangle} \leq sup_{x \in A} {\langle u, x \rangle} $ and $ {\langle u, b \rangle} \leq sup_{y \in B} {\langle u, y \rangle} $ so for compatibility reasons with the formula of decomposition of $c$, the inequalities must be equalities. It gives $ {\langle u, a \rangle} = sup_{x \in A} {\langle u, x \rangle} $ and $ {\langle u, b \rangle} = sup_{y \in B} {\langle u, y \rangle} $ which means $ u \in C_D(a) \cap C_D(b) $.

Now we have to ask ourselves what are the conditions to get a Minkowski vertex when we compute the intersection $C_D(a) \cap C_D(b)$?

Let $A$ and $B$ be $\mathbb{R}^n$-polytopes of full dimension $n$, $\mathcal{V}_A$ and $\mathcal{V}_B$ be their vertices lists. Let $ a \in \mathcal{V}_A $, and $ b \in \mathcal{V}_B $:

\begin{equation}
  c=(a+b) \in \mathcal{V}_C \Leftrightarrow dim( C_D(a) \cap C_D(b) )=n.
\label{dim_cone_sommet}
\end{equation}

In \cite{Fukuda2005_882}, Fukuda and Weibel indicate that \textit{``Faces of a polytope and their normal cones have dual properties. In particular, if F is a $i$-dimensional face of A, then the normal cone $C_D(F)$ is a $( n-i )$-dimensional cone of $\mathbb{R}^n$.''} So $ \exists a \in \mathcal{V}_A, \exists b \in \mathcal{V}_B / c=(a+b)  \in \mathcal{V}_C \Rightarrow dim( C_D(a) \cap C_D(b) )=n $ as $c$ is a $0$-face.

Reciprocally if $ dim( C_D(a) \cap C_D(b) )=n $ then $ \exists u \in Int( C_D(a) ) \cap Int( C_D(b) ) $ such as $ \forall x \in A, {\langle u, a \rangle} > {\langle u, x \rangle} $ and $ \forall y \in B, {\langle u, b \rangle} > {\langle u, y \rangle} $. So $ {\langle u, a \rangle} + {\langle u, b \rangle} = {\langle u, c \rangle} > {\langle u, x+y \rangle}, \forall (x+y) \in C $. Hence $ dim( C_D(a) \cap C_D(b) )=n \Rightarrow c \in \mathcal{V}_C $.

Following the same idea we can find the facets of the polytope $C$ from $N(C)$ edges. We now have all the tools to build Minkowski vertices and facets.

\subsection{A first dual algorithm}

\begin{algorithm}   
\caption{Calculate $C = A+B$ with $A$ and $B$, two $\mathbb{R}^n$-polytopes of full dimension $n$}
\label{algbrut}     
\begin{algorithmic}
\REQUIRE List of dual cones of A $\{ C_D(a_i), a_i \in \mathcal{V}_A \}$
\REQUIRE List of dual cones of B $\{ C_D(b_j), b_j \in \mathcal{V}_B \}$
\FORALL{$a_i \in \mathcal{V}_A$ and $b_j \in \mathcal{V}_B$}
	\STATE Compute $C_D(a_i) \cap C_D(b_j)$
	\IF {$ dim( C_D(a_i) \cap C_D(b_j) ) = n $}
		\STATE $ (a_i+b_j) \in \mathcal{V}_C $
		\STATE Get the half-spaces passing through $(a_i+b_j)$ from $C_D(a_i) \cap C_D(b_j)$ edges
	\ENDIF
\ENDFOR
\end{algorithmic}
\end{algorithm}

\subsection{Complexity and implementation}

To perform such an operation, in \cite{Fukuda96doubledescription} Fukuda gives interesting insights and efficient-to-use strategies despite the fact, as the author says, \textit{"that we can hardly state any interesting theorems on its time and space complexities"}. The underlying physical problem we designed this algorithm for, is in low dimension so obtaining the polytopes double description is not a problem. In tolerancing analysis, it can even be done in an analytical way but beware that the number of vertices and facets can be exponential according to the dimension of the space we work in. As a example a tetraedon in $\mathbb{R}^n$ has only $(n+1)$ vertices and $(n+1)$ facets but a cube has $2n$ facets and $2^n$ vertices. In \cite{Chazelle1993} one can find an optimum algorithm, when $n$ is constant, to compute convex hulls that runs in time $\mathcal{O}(k^{\lceil n/2 \rceil})$ for $n \ge 4$, $k$ being the number of vertices. Such an upper bound cannot be reduced as it is of the order of the larger output. So in high dimensions the kind of polytopes you handle can have a very strong impact on the performances of the Minkowski sum algorithm. In \cite{Avis1997265} we can find a good introduction on families of polytopes and the way they behave in algorithmic contexts.

This algorithm has been coded in C++ and is available under the Gnu General Public Licence v3.0 at \url{http://i2m.u-bordeaux.fr/politopix}. It has been tested and is now fully operational in $\mathbb{R}^6$ in the frame of tolerance analysis in mechanical engineering, as well as it is in any dimension given the limitations we previously described.

\subsection{An optimized dual algorithm}

The basic idea behind this algorithm is quite simple. As soon as we get $C_D(a_i)$ and $C_D(b_j)$ such as $ dim( C_D(a_i) \cap C_D(b_j) )=n$, we do not pick dual cones from $A$ in a random way but we rather select the list of $C_D(a_i)$ neighbours to intersect them with $C_D(b_j)$. While we find intersections of dimension $n$, we keep on picking up the neighbours of the neighbours and so on.

Let's assume that the two dual cones $C_D(a_i)$ and $C_D(b_j)$ intersects with each other such as, at least one half-space hyperplane $\bar{H}$ of $C_D(a_i)$ separates $C_D(b_j)$. Then it is obvious that the neighbour dual cone $C_D(a_k)$ that shares $\bar{H}$ with $C_D(a_i)$ has also a non empty intersection with $C_D(b_j)$. We can take advantage of this neighbourdhood property to speed up the algorithm.

So we introduce the notion of polyhedral cap. Let $A$, $B$ and $C=A+B$ be $\mathbb{R}^n$-polytopes and $\mathcal{V}_A$, $\mathcal{V}_B$, $\mathcal{V}_C$ their respective lists of vertices. For a given vertex $a \in \mathcal{V}_A$ we want to find the list of all the vertices of $B$ that will give a Minkowski vertex in $C$. We define the polyhedral cap of the vertex  $a_i$ in the polytope $B$ this way $ \mathcal{V}_B^+(a_i) = \{ b_j \in \mathcal{V}_B / (a_i+b_j) \in \mathcal{V}_C \} $, its complementary list in $\mathcal{V}_B$ is  $ \mathcal{V}_B^-(a_i) = \{ b_j \in \mathcal{V}_B / (a_i+b_j) \notin \mathcal{V}_C \}$ .

Let $A$ et $B$ be two $\mathbb{R}^n$-polytopes,  $a_i \in \mathcal{V}_A$ :

\begin{equation}
\forall i, \mathcal{V}_B= \mathcal{V}_B^+(a_i) \cup \mathcal{V}_B^-(a_i)
\end{equation}
\begin{equation}
\forall i, \mathcal{V}_B^+(a_i) \not= \emptyset, \mathcal{V}_B^+(a_i) \text{ is connected}.
\end{equation}

We want to show that $ \forall i, \mathcal{V}_B^+(a_i) \neq \emptyset $. By definition $ a_i \in \mathcal{V}_A $ so $ dim(C_D(a_i)) = n $, let $l$ be $ Card(\mathcal{V}_B) $, so $ \forall j \in \{1, ..., l \} $ we have $ dim(C_D(b_j)) = n $. Let's admit that for the $l-1$ first dual cones of $B$ we have $ \forall j \in \{1, ..., l-1 \} $ $ dim(C_D(a_i) \cap C_D(b_j)) < n $, which means that they do not intersect with $C_D(a_i)$ or only with its frontier. Given that $ \bigcup_{j=1}^l C_D(b_j) = \mathbb{R}^n $ it must only be the last dual cone in $N(B)$ that intersects with $C_D(a_i)$ interior. As a consequence $ dim(C_D(a_i) \cap C_D(b_l)) = n $ so $ a_i + b_l \in \mathcal{V}_C $. In all the cases we have at least one vertex in $B$ that verifies this property so $ \mathcal{V}_B^+(a_i) \neq \emptyset $.

\begin{algorithm}   
\caption{Calculate $C = A+B$ with $A$ and $B$, two $\mathbb{R}^n$-polytopes of full dimension $n$}
\label{algopt}     
\begin{algorithmic}
\REQUIRE List of dual cones of A $\{ C_D(a_i), a_i \in \mathcal{V}_A \}$
\REQUIRE List of dual cones of B $\{ C_D(b_j), b_j \in \mathcal{V}_B \}$
\FORALL{$a_i \in \mathcal{V}_A$}
	\STATE $ findMinkVertex = false $
	\WHILE{$b_j \in \mathcal{V}_B$ and $ findMinkVertex==false $}
		\STATE Compute $ C_D(c_{ij}) = C_D(a_i) \cap C_D(b_j) $
		\IF {$ dim( C_D(c_{ij}) ) = n $}
			\STATE $ findMinkVertex = true $
			\STATE $ c_{ij}=(a_i+b_j) \in \mathcal{V}_C $
			\STATE \textit{// Now find $a_i$ polyhedral cap.}
			\STATE ProcessNeighbours($C_D(a_i), C_D(b_j), C_D(c_{ij})$)
		\ENDIF
	\ENDWHILE
\ENDFOR

\end{algorithmic}
\end{algorithm}

We shall proove now that $ \mathcal{V}_B^+(a) $ is connected. Let $ c_1=a+b_1 \in \mathcal{V}_C $ and $ c_2=a+b_2 \in \mathcal{V}_C $ i.e. $ b_1 \in \mathcal{V}_B^+(a) $ and $ b_2 \in \mathcal{V}_B^+(a) $. Is there a path of vertices of $B$, leading from $b_1$ to $b_2$ in $\mathcal{V}_B^+(a)$? We know that $ C_D(c_1) = C_D(a) \cap C_D(b_1) $ and $ C_D(c_2) = C_D(a) \cap C_D(b_2) $. We choose in the interiors $ u_1 \in Int( C_D(a) \cap C_D(b_1) ) $ and $ u_2 \in Int( C_D(a) \cap C_D(b_2) ) $ so $ [u_1,u_2] \subset Int( C_D(a) ) $. We build the list $L$ of dual cones in $N(B)$ whose intersection is not empty with the segment $[u_1,u_2]$ and that verify the following property: once we leave one of the bounding half-spaces of the current cone $C_D(b_j)$, we add to $L$ its neighbour cone that shares with $C_D(b_j)$ the frontier of this current half-space. We can note that once we leave a cone, we will never process it again, as we remain on the segment $[u_1,u_2]$ we can never re-enter a bounding half-space we previously left. So $L$ is a finite ordered list. We can also say that if $[u_1,u_2]$ passes through the interior of $C_D(b_j)$ then $ (a+b_j) $ is a Minkowski vertex in $C$. On the other side if $[u_1,u_2]$ intersects $C_D(b_j)$ only with its frontier then as $ [u_1,u_2] \subset Int( C_D(a) ) $ it is easy to build a point still inside $C_D(a)$ that will also be in the interior of $C_D(b_j)$, we only have to shift away from $[u_1,u_2]$ by a very small quantity. In all the cases $ (a+b_j) \in \mathcal{V}_C $, moreover two consecutive dual cones in $L$ are connected, so are their corresponding vertices in $B$. So $L$ is a list having for first and last elements $b_1$ and $b_2$ connected through neighbour vertices $ \{ b_j, b_j \in \mathcal{V}_B \} $ such that $ (a+b_j) \in \mathcal{V}_C $. So $ \mathcal{V}_B^+(a) $ is connected.

\begin{algorithm}
\begin{algorithmic}
\STATE \textit{// Find neighbours through common facets}

PROCEDURE
\STATE 
ProcessNeighbours($C_D(a_i), C_D(b_j), C_D(c_{ij})$)
	\STATE Mark $C_D(b_j)$ as processed
	\FORALL{$ C_D(b_k) \in N(B)$, $F_l$ facet of $C_D(c_{ij})$ such that $ F_l = C_D(b_k) \cap C_D(c_{ij}) $ }
		\STATE \textit{// At this step $b_j$ and $b_k$ are neighbours in B}
		\IF {$C_D(b_k)$ is not marked as processed}
			\STATE Compute $ C_D(c_{ik}) = C_D(a_i) \cap C_D(b_k) $
			\IF {$ dim( C_D(c_{ik}) ) = n $}
				\STATE $ c_{ik}=(a_i+b_k) \in \mathcal{V}_C $
				\STATE ProcessNeighbours($C_D(a_i), C_D(b_k), C_D(c_{ik})$)
			\ENDIF
		\ENDIF
	\ENDFOR

END PROCEDURE

\end{algorithmic}
\end{algorithm}

\section{Minkowski sum of polytopes with primal cones}

We will now consider the primal cones of the polytopes we handle that is to say, for each vertex $ a \in \mathcal{V}_A $, the cones defined by the intersection of the half-spaces $ \{ \bar{H}_u^+(a) \} $ attached to this current vertex. In an equivalent manner, we can say that, given $ a \in \mathcal{V}_A $, such cones can be generated by all the edges attached to $a$ and pointing towards its neighbour vertices $ \{ a_i \} $. Let's write $\mathcal{V}_A(a)$ the set of the vertices of $\mathcal{V}_A$ adjacent to $a$, with $\alpha_i \in \mathbb{R}^+, a_i \in \mathcal{V}_A(a)$

\begin{equation}
C(a) = a + \sum_{i} \alpha_i (a_i-a) = \bigcap_{u} \bar{H}_u^+(a).
\end{equation}

\begin{figure*}
  \begin{center}
    \includegraphics[scale=0.66]{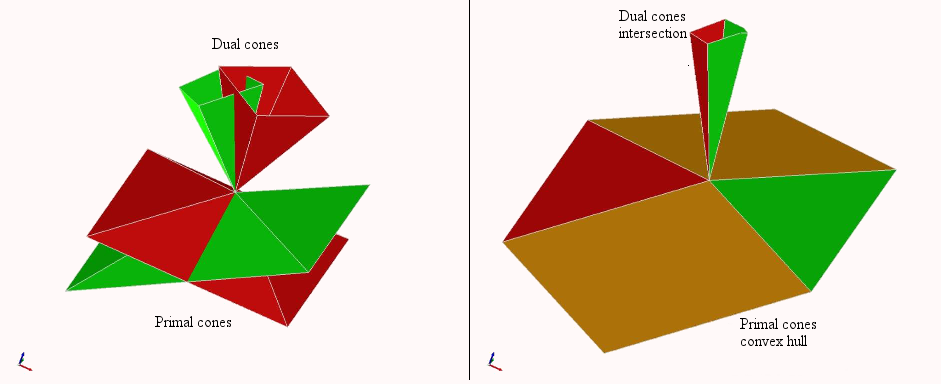}
    \caption{\fontsize{8}{9.6}{On the left two primal cones on the bottom with their duals on top of the drawing, on the right the primal cones convex hull and the duals intersection.}}
    \label{inter_hull}
  \end{center}
\end{figure*}

\subsection{Main properties}

Let $C_1$ and $C_2$ be two cones attached to the origin:
\begin{equation}
C_1 + C_2 = Conv(C_1, C_2).
\end{equation}

Let $x \in C_1 + C_2$ then $ \exists y \in C_1, \exists z \in C_2 $ such as $x = y+z = \frac{1}{2} 2y + \frac{1}{2} 2z$. Yet $ 2y \in C_1$ and $ 2z \in C_2$ which means that any point in $C_1 + C_2$ can be written as a point included in the convex hull of $C_1$ and $C_2$ hence $ C_1 + C_2 \subset Conv(C_1, C_2) $.

Let $x \in Conv(C_1, C_2)$ then $ \exists y \in C_1, \exists z \in C_2 $ and $\theta \in [0,1]$ such as $x = \theta y + (1-\theta) z$. Yet $ \theta \geq 0 $ so $ \theta y \in C_1 $,  $ 1-\theta \geq 0$ hence $ (1 - \theta) z \in C_2 $ which means that $x$ can be written as the sum of two elements from $A$ and $B$ so $ Conv(C_1, C_2) \subset C_1 + C_2 $.

It is easy to transpose this property to the case of two cones attached to the vertices $a$ and $b$ in the polytopes $A$ and $B$ provided that we translate them first in $a+b$ and hereafter compute the convex hull:

\begin{equation}
C(a) + C(b) = Conv \Big( b + C(a), a + C(b) \Big).
\end{equation}

If $c = a+b$ is a Minkowski vertex of $C = A+B$ then:

\begin{equation}
C(c) = Conv \Big( b + C(a), a + C(b) \Big).
\label{basic_primal}
\end{equation}

From \cite{Fukuda20041261} if $c$ and $c'$ are two adjacent vertices in $C$ with their given decomposition in elements of $A$ and $B$, $ c=a+b $ et $ c'=a'+b' $ then $a$ et $a'$ are either equal or adjacent (respectively $b$ and $b'$). We deduce that the list of edges defining $C(c)$ is a sublist of $ L(C(a), C(b)) $, defined as the list of edges of both $C(a)$ and $C(b)$ translated in $c$. So $ C(c) \subset Conv(b+C(a), a+C(b)) $ because $ Conv(b+C(a), a+C(b)) $ is entirely defined by $ L(C(a), C(b)) $.

For the reciprocal we use the dual, we know that $ C_D(c) = C_D(a) \cap C_D(b) $ so
\[
\begin{cases}
C_D(c) \subset C_D(a) \\
C_D(c) \subset C_D(b).
\end{cases}
\]

These relations are reversed in the primal space:
\[
\begin{cases}
b+C(a) \subset C(c) \\
a+C(b) \subset C(c).
\end{cases}
\]

Hence $ Conv(b+C(a), a+C(b)) \subset C(c) $ because $C(c)$ is convex. Now we can give the property linking primal and dual cones respectively attached to the vertices $ a_i \in \mathcal{V}_A $ and $ b_j \in \mathcal{V}_B $ when their sum provide a Minkowski vertex:

$ c \in \mathcal{V}_C \Rightarrow $
\begin{equation}
\Big( C_D(a) \cap C_D(b) \Big)_D = Conv \Big( b + C(a), a + C(b) \Big).
\end{equation}

The proof is quite straightforward, as $c$ is a Minkowski vertex then $ C_D(c) = C_D(a) \cap C_D(b) $ and the dual of the dual is the primal so $ ( C_D(c) )_D = C(c) = Conv(b + C(a), a + C(b)) $.

This property is fundamental in the sense that it can make the connection between a polyhedra intersection problem on one side, and a polyhedra convex hull computation on the other. In the context of the sums of polytopes we're aware that if $c = a+b$ is a Minkowski vertex of $C = A+B$ then computing the convex hull of the two primal cones $C(a)$ and $C(b)$ is equivalent to computing the intersection between their corresponding duals $C_D(a)$ and $C_D(b)$, see Fig. \ref{inter_hull}. As a consequence we can compute $ C(c) $ $ \forall c \in \mathcal{V}_C $ - which means we can find all facets of $C$ - with data coming only from the primal space. Given that a polytope is entirely determined by its vertices or facets we can write the following property for $A$ and $B$ two $\mathbb{R}^n$-polytopes with respectively $k$ and $l$ vertices. Let's note $ K $ the set of indices that provide a Minkowski vertex in $ C=A+B $ i.e. $ (i,j) \in K_{A+B} \Leftrightarrow (a_i+b_j) \in \mathcal{V}_C $.

\begin{equation}
  A+B = \bigcap_{i=1}^k C(a_i) + \bigcap_{j=1}^l C(b_j) = \bigcap_{(i,j) \in K} C(a_i) + C(b_j)
\end{equation}

We can generalize this property to all the sums of vertices as it is easy to proove that whatever $ a_i \in \mathcal{V}_A $ and $ b_j \in \mathcal{V}_B $, $ A+B \subset C(a_i) + C(b_j) $ because we know that $ A \subset C(a_i) $ and $ B \subset C(b_j) $. So it is clear that:

\begin{equation}
  A+B = \bigcap_{i=1,j=1}^{k,l} \Big( C(a_i) + C(b_j) \Big).
  \label{primal_cones_sum}
\end{equation}

\subsection{A primal algorithm}

With the property \ref{basic_primal}, it is quite easy to set up an algorithm computing the polytope $C=A+B$.

\begin{algorithm}   
\caption{Calculate $C = A+B$ with $A$ and $B$, two $\mathbb{R}^n$-polytopes of full dimension $n$}
\label{algopt}     
\begin{algorithmic}
\REQUIRE List of primal cones of A $\{ C(a_i), a_i \in \mathcal{V}_A \}$
\REQUIRE List of primal cones of B $\{ C(b_j), b_j \in \mathcal{V}_B \}$
\REQUIRE $ a_u \in \mathcal{V}_A $ and $ b_v \in \mathcal{V}_B $ such as $ a_u+b_v \in \mathcal{V}_C $

\STATE ProcessCones($C(a_u), C(b_v)$)

\end{algorithmic}
\end{algorithm}
 
\begin{algorithm}
\begin{algorithmic}
\STATE ~

PROCEDURE
\STATE 
ProcessCones($C(a_i), C(b_j)$)
	\STATE Mark $ c_{ij}=(a_i+b_j) \in \mathcal{V}_C $ as processed
	\STATE Find facets, edges of $ C(c_{ij})= Conv(C(a_i), C(b_j)) $ 
	\FORALL{edges of $ C(c_{ij}) $ leading to $ c_{uv} \in \mathcal{V}_C $ }
		\IF {$c_{uv}$ is not marked as processed}
			\STATE ProcessCones($C(a_u), C(b_v)$)
		\ENDIF
	\ENDFOR
	
END PROCEDURE
\end{algorithmic}
\end{algorithm}

Starting from a first Minkowski vertex $c_1=a_1+b_1$ we just need to compute its edges making sure they belong to the convex hull of the cone $C(c_1)$. Following such edges will leed to $c_1$ neighbours where we will compute their corresponding convex hulls. At this step it is important to note that, to identify Minkowski vertices, if $c_2=a_2+b_2$ is a neighbour if $c_1$ then $a_1$ and $a_2$ are either equal or adjacent in $A$, $b_1$ and $b_2$ are either equal or adjacent in $B$. The edges of $C$ are either parallel to an edge of $A$ or an edge of $B$.

\subsection{Complexity}

We have reduced our problem to a convex hull algorithm. However as stated in \cite{Fukuda2004} \textit{``we are still very far from knowing the best ways to compute the convex hull for general dimensions''} and the author adds \textit{``in the general case, there is no known polynomial algorithm''}. Despite the current state-of-the-art we believe we do not have only a theoritical achievement with property \ref{primal_cones_sum}. The reason is due to the fact that it is not just about computing the convex hull of a set of edges coming from cones $C_1$ and $C_2$ but rather computing the convex hull of two sets of edges, each one of them being already convex. As building a convex set from two convex sets is easier, we plan to explore this track in the future.

\section{Conclusion}

We have developped and implemented a dual algorithm based on the intersection of dual cones used in the field of tolerance analysis where it behaves very well in terms of robustness and computation time. However the fact that we handle the double description to sum polytopes could possibly be a limitation if one needs to work in high dimensions. In the second part, we introduced new properties to remain in the primal space and proposed another way to perform the operation. Now we plan to work on a parallel version of the first as well as improving the theoritical background of the second.

\end{document}